\documentclass[twocolumn,twoside,slac_two]{revtex4}
\usepackage{graphicx}
\usepackage{fancyhdr}
\usepackage{graphics}
\usepackage{epstopdf}
\usepackage{textpos}
\usepackage{pgf}
\begin{document}

\title{\centering $\bf{W}$/$\bf{Z}$ and Diboson Properties}


\author{
\centering
\includegraphics[scale=0.15]{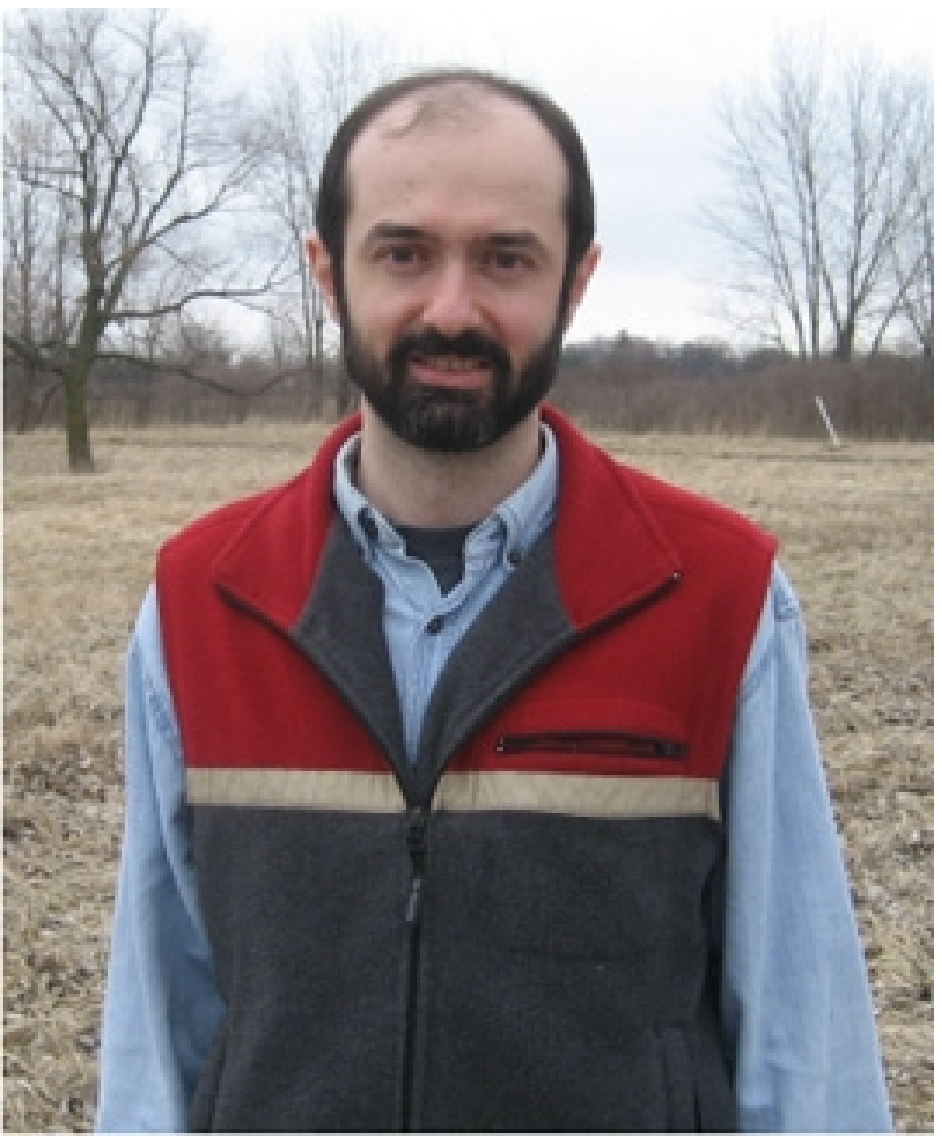} \\
\begin{center}
Alex Melnitchouk for ATLAS, CDF, CMS, D0, and LHCb Collaborations
\end{center}}
\affiliation{\centering University of Mississippi, University, Mississippi, 38677, USA}
\begin{abstract}
Most recent measurements of properties of $W$ and $Z$  gauge bosons at hadron colliders
are presented. The measurements were performed by ATLAS, CMS, and LHCb collaborations with proton-proton collisions at the LHC and by
CDF and D0 collaborations with proton-antiproton collisions at the Tevatron. 
Center-of-mass energy was 7 TeV and 1.96 TeV at the LHC and Tevatron respectively.
Integrated luminosity ranges from 35 pb$^{-1}$ to 1.02 pb$^{-1}$ for LHC and
from 0.2 fb$^{-1}$ to 7.1 fb$^{-1}$ for Tevatron.
\end{abstract}

\maketitle
\thispagestyle{fancy}


\section{INTRODUCTION}
With large data samples containing $W$ and $Z$ bosons (e.g. millions of $W$ events at the Tevatron experiments) that are currently available for analysis,
precise measurements of electroweak properties became possible. At the same time, the LHC experiments
after successful LHC commissioning and start-up are already peforming electroweak measurements
with rapidly increasing data samples.
In this paper we divide $W$, $Z$ and diboson properties in three groups:
\begin{enumerate}
\item Couplings between electroweak gauge bosons.
\item $Z$ boson properties.
\item $W$ boson properties.
\end{enumerate}
The paper consists of three main sections that reflect this classification.
Typical event selection in such measurements include one or more high $p_T$ isolated leptons\footnote{electrons or muons in the context of
this usage}.
Main backgrounds include electroweak processes other than the process of interest (e.g. $Z \to ee$ can be a background to $W \to e\nu$),
QCD processes in which a quark or a gluon jet is mis-identified for an isolated lepton, combination of the two mentioned backgrounds 
(e.g. $Z$+jets can be a background to $WZ$).

\section{COUPLINGS BETWEEN ELECTROWEAK GAUGE BOSONS}
Triple Gauge Couplings (TGCs) for the following vertices are considered: 
\begin{enumerate}
\item $W W V$,
\item $Z \gamma V$,
\item $Z Z V$,
\end{enumerate}
where $V$ is $Z$ or $\gamma$.
$W  WV$ vertex involves {\it charged} couplings (charged TGCs) whereas  $Z \gamma V$ and $Z Z V$ involve
{\it neutral} couplings (neutral TGCs). 

\subsection{Charged TGCs}
Most general effective Lagrangian on which independent conservation of C and P is imposed contains five coupling parameters:
$g_1^Z$, $K_{\gamma}$, $K_Z$, $\lambda_{\gamma}$, $\lambda_Z$. 

\begin{figure}
\includegraphics[width=75mm]{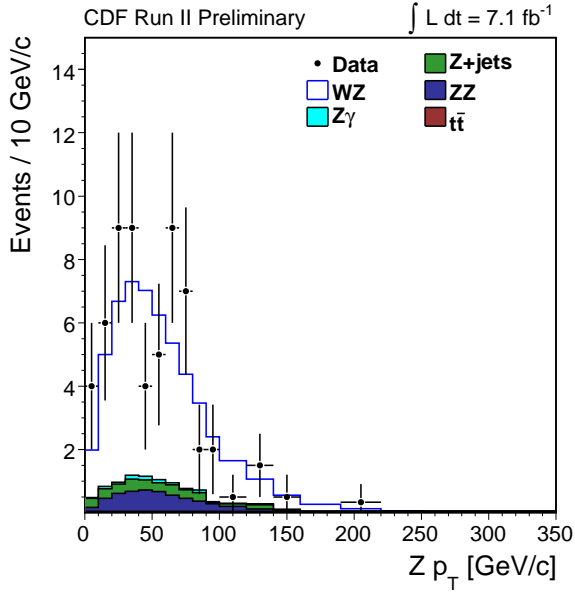}
\caption{$Z$ boson $p_T$ distribution in three leptons plus missing transverse energy (CDF).}
\label{fig:cdfzptinwz}
\end{figure}

\begin{figure}
\includegraphics[width=75mm]{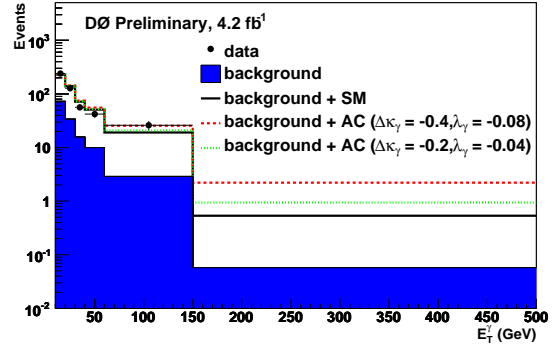}
\caption{Photon transverse energy in $W \gamma$ events (D0).}
\label{fig:d0gammaptinwgamma}
\end{figure}

In the Standard Model (SM) $g_1^Z$,$K_{\gamma}$, $K_Z$ are all equal to one and $\lambda_{\gamma}$, $\lambda_Z$ are
both equal to zero. Deviations of these couplings from their predicted SM values are denoted with preceding $\Delta$, 
e.g. $\Delta K_{\gamma}$ = $K_{\gamma}$ - 1. Two of the five couplings are linked with the W boson quadrupole electric moment $Q_W$ and magnetic
dipole moment $\mu_W$:
\begin{eqnarray*}
  Q_W & = & - {e \over M_W^2} ({1 + \Delta K_{\gamma} - \lambda_{\gamma}})\\
  \mu_W & = & - {e \over M_W^2} ({2 + \Delta K_{\gamma} + \lambda_{\gamma}})
\end{eqnarray*}
In a more general theory $g_1^Z$, $K_{\gamma}$, $K_Z$, $\lambda_{\gamma}$, $\lambda_Z$ can deviate from their SM values 
(anomalous couplings).
Such deviations would violate unitarity and therefore must be regulated with the following form factor 
\begin{eqnarray*}
  \alpha(\hat{s}) & = & {\alpha_0} \over {(1 + \hat{s}/\Lambda^2)}^2
\end{eqnarray*}
where $\alpha(\hat{s})$ is any of the five anomalous couplings for center-of-mass partonic energy $\sqrt{\hat{s}}$,
and $\alpha_0$ is its low-energy approximation; $\Lambda$ is energy scale of physics beyond SM.
In this description, the anomalous couplings vanish as $\hat{s} \rightarrow \infty$, hence unitarity is restored.
Charged TGCs contribute to $W W$, $W Z$, and $W \gamma$ final states.  
Anomalous values of charged TGCs would lead to enhanced production rate and more energetic distributions of final state gauge bosons, and,
consequently, more energetic distributions of their decay products relative to SM case. 
Most stringent limits on
$g_1^Z$, $\lambda_Z$, $K_Z$ are set by CDF~\cite{cdfwz} with 7.1fb$^{-1}$ using $W Z$ final state with three leptons and missing transverse energy.
Transverse momentum of the $Z$ boson shown in Figure~\ref{fig:cdfzptinwz} is used to set limits.
One-dimensional limits are shown in Table~\ref{tab:cdfwzlimits}.
\begin{table}[t]
\begin{center}
\caption{CDF limits on  $\lambda_Z$, $g_1^Z$, $K_Z$ at 7.1fb$^{-1}$. No relationship between couplings is assumed.}
\begin{tabular}{|l|c|c|c|}
\hline \textbf{ } & \textbf{$\lambda_Z$} & \textbf{$\Delta g_1^Z$} &
\textbf{$\Delta K_Z$}
\\
\hline 1.5TeV & -0.08 - 0.10 & -0.09 - 0.22 & -0.42 - 0.99 \\
       2.0TeV & -0.09 - 0.11 & -0.08 - 0.20 & -0.39 - 0.90 \\
\hline
\end{tabular}
\label{tab:cdfwzlimits}
\end{center}
\end{table}
Most stringent limits on
$K_{\gamma}$ and $\lambda_{\gamma}$ from hadron collider are set by D0~\cite{d0wgamma} with 4.2fb$^{-1}$ using $W \gamma$ in the muon channel.
Photon transverse energy shown in Figure~\ref{fig:d0gammaptinwgamma} is used to set limits.
One-dimensional limits are shown in Table~\ref{tab:d0wgammalimits}
\begin{table}[t]
\begin{center}
\caption{D0 limits on $K_{\gamma}$ and $\lambda_{\gamma}$. No relationship between couplings is assumed.}
\begin{tabular}{|l|c|c|c|}
\hline \textbf{ } & \textbf{$\Delta K_{\gamma}$} & \textbf{$\Delta \lambda_{\gamma}$} 
\\
\hline 2.0TeV & -0.14 - 0.51 & -0.12 - 0.13  \\
\hline
\end{tabular}
\label{tab:d0wgammalimits}
\end{center}
\end{table}

\subsection{Neutral TGCs}
Similarly to the case of charged TGCs neutral TGCs appear in the effective Langrangian.
In case of $Z \gamma V$ vertex these are $h_1^V$, $h_2^V$, $h_3^V$, $h_4^V$ couplings.
In case of $Z Z V$ vertex these are CP-violating $f_4^V$ and CP-conserving $f_5^V$ couplings.

\subsubsection{$Z \gamma V$ vertex}
\begin{figure}
\includegraphics[width=75mm]{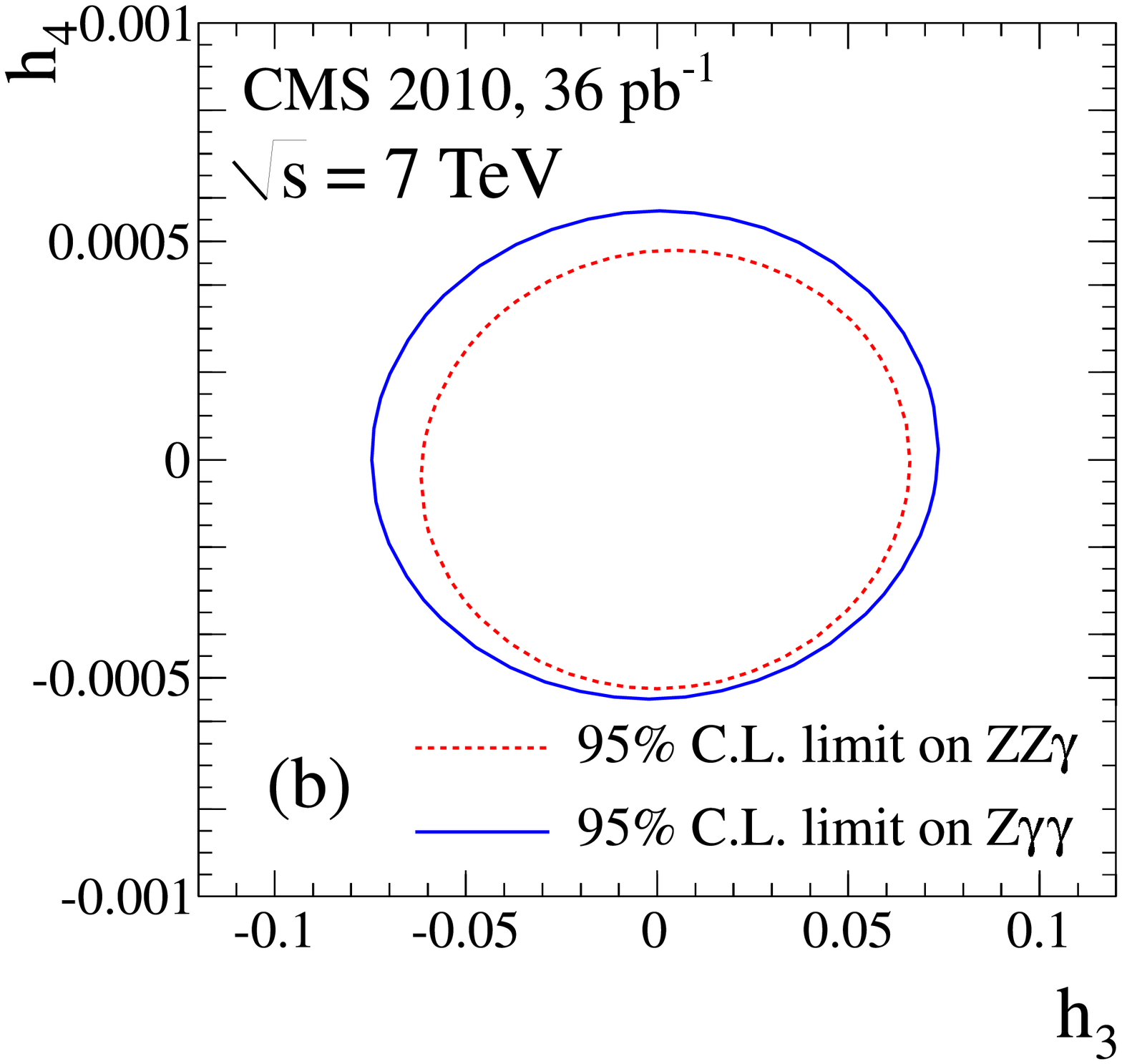}
\includegraphics[width=75mm]{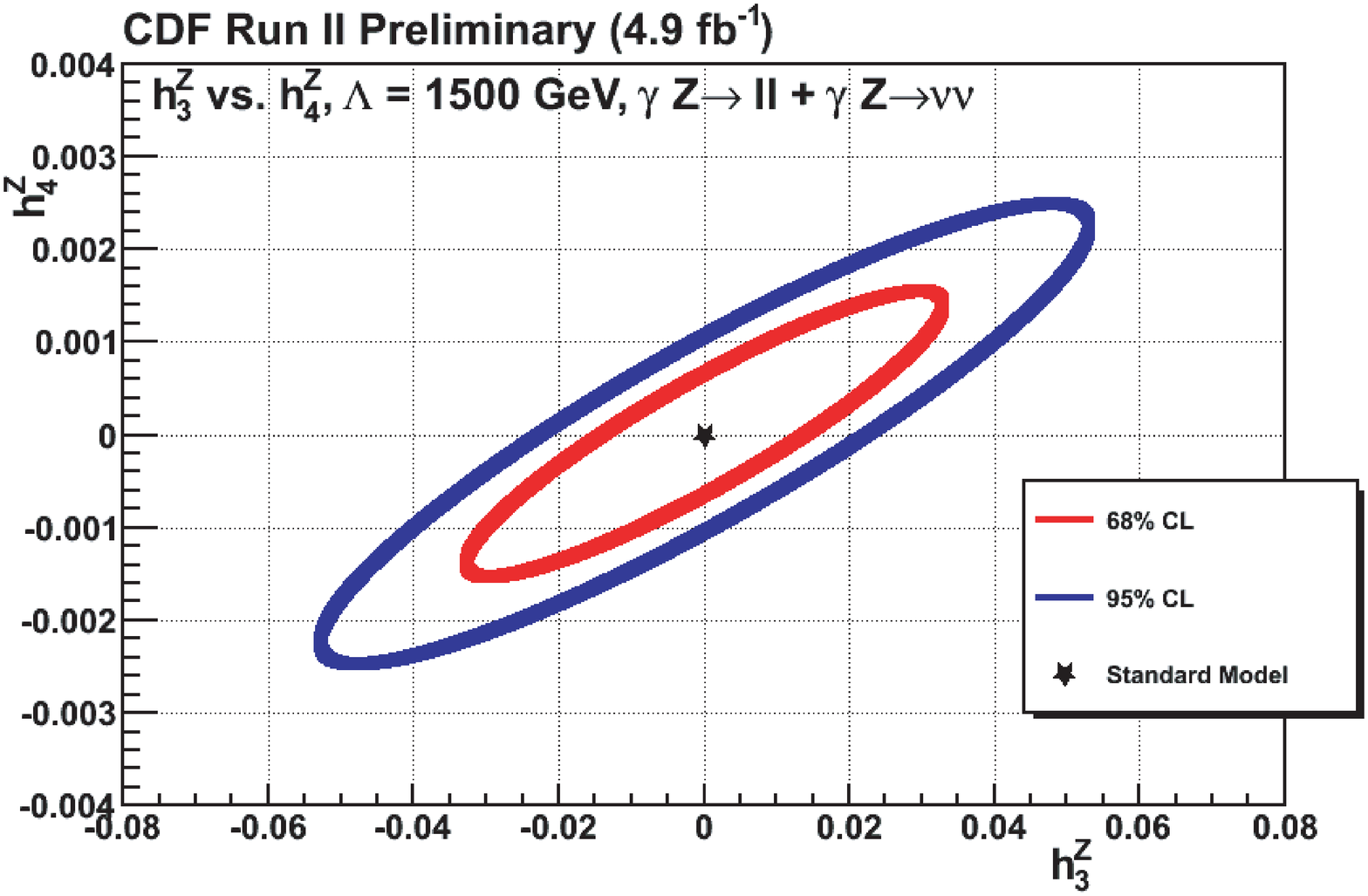}
\includegraphics[width=75mm]{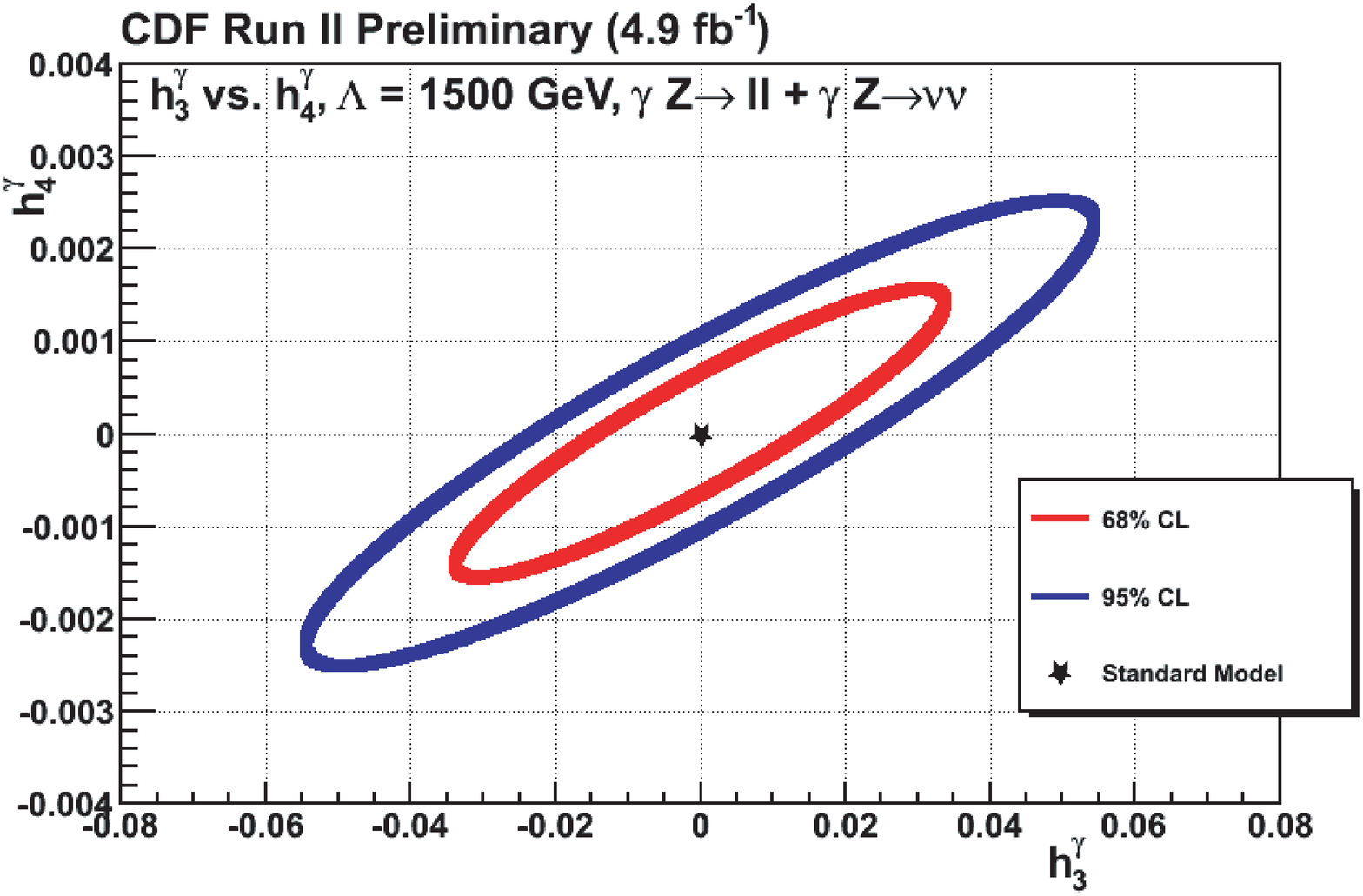}
\caption{Limits on $h_3^V$ and $h_4^V$ neutral TGCs}
\label{fig:zgammalimits}
\end{figure}

We report results on CP-conserving couplings probed with $Z \gamma$ final state.
Two-dimensional limits from CMS~\cite{cmsewk10008} and CDF~\cite{cdfprl107}, obtained with 36pb$^{-1}$ and 4.9fb$^{-1}$ respectively, 
are shown in Figure~\ref{fig:zgammalimits}

\subsubsection{$Z \gamma V$ vertex}
\begin{figure}
\includegraphics[width=75mm]{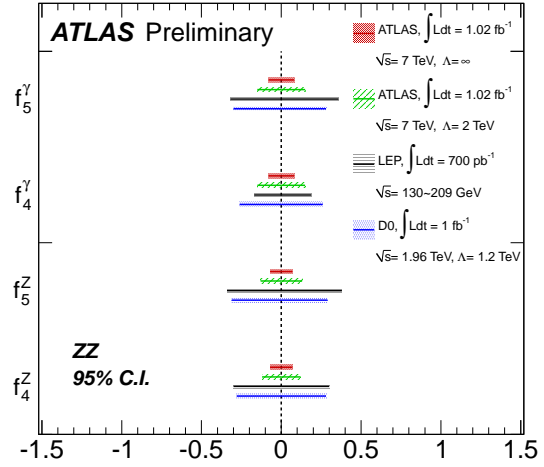}
\caption{Limits on $f_4^V$ and $f_5^V$ neutral TGCs}
\label{fig:atlaszz}
\end{figure}

We report results on both CP-violating $f_4^V$ and CP-conserving $f_5^V$ couplings probed with $Z Z$ final state.
Most stringent one-dimensional limits from ATLAS~\cite{ATLAS_CONF_2011_107} obtained with 1.02fb$^{-1}$ together with comparisons 
are shown in Figure~\ref{fig:atlaszz}.

\section{$\bf{Z}$ BOSON PROPERTIES}
Measurements of the following $Z$ boson properties are presented:
\begin{enumerate}
\item Forward-backward asymmetry.
\item Effective weak mixing angle.
\item Couplings between $Z$ boson and fermions.
\item Angular coefficients.
\end{enumerate}

\subsection{Forward-backward asymmetry}
Forward-backward asymmetry is a consequence of presence of two types of couplings 
between $Z$ boson and final state fermions (vector and vector-axial). 
Asymmetry is defined as a ratio of difference and sum of forward and backward events.
Event is classified as forward (backward) if $cos\theta*$ is positive (negative).
Angle $\theta*$ is scattering angle measured in Collins-Soper reference frame~\cite{csframe}.
The sign and degree of asymmetry depends on relative content of $Z$, $\gamma$, and interference processes.
Therefore it depends on the mass of the $Z$ boson. Degree of asymmetry also depends on relative content
of $u\bar{u}$ versus $d\bar{d}$ collisions. Hence forward-backward asymmetry is sensitive to weak mixing angle and couplings between 
$Z$ boson and fermions.

\begin{figure}
\includegraphics[width=75mm]{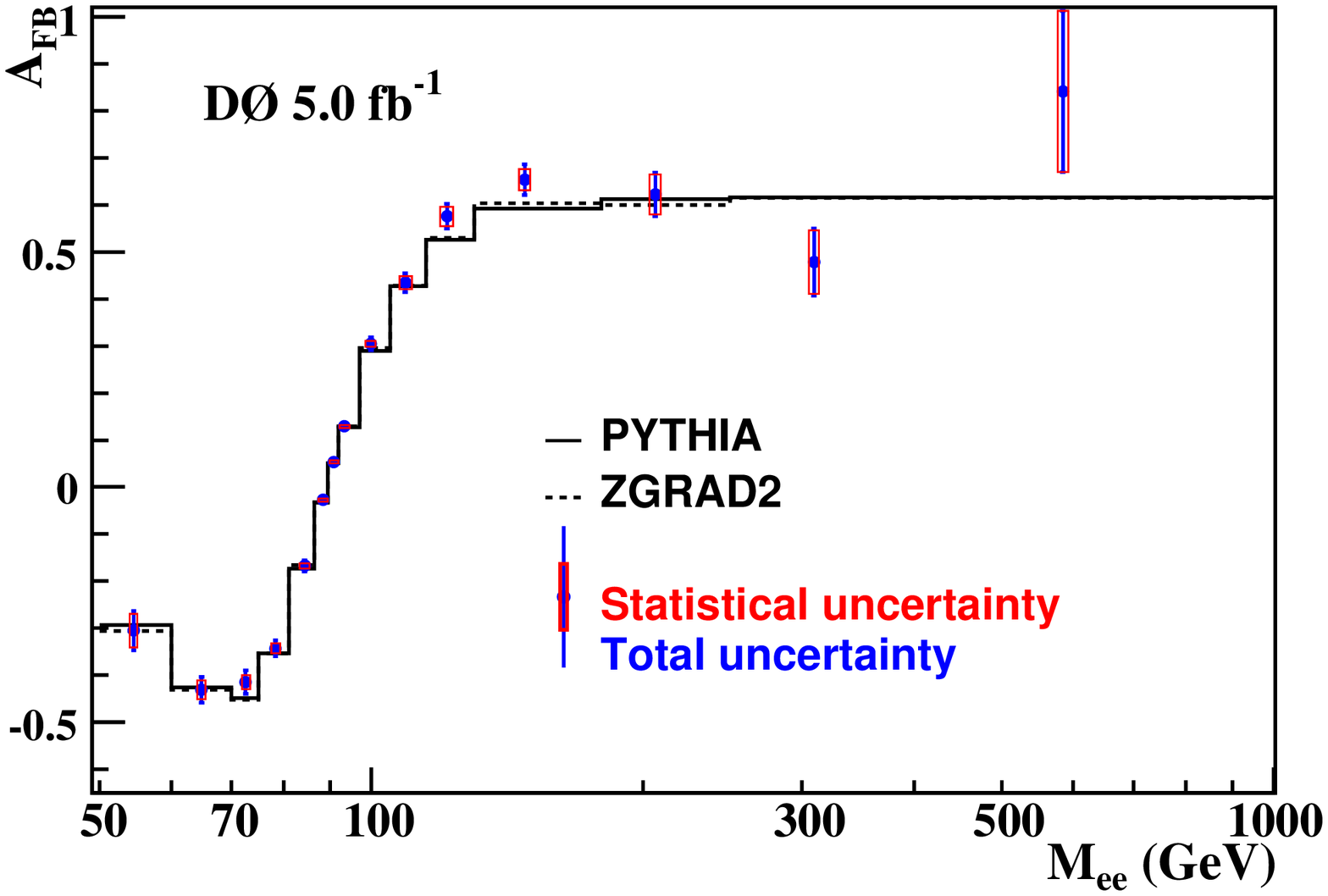}
\includegraphics[width=75mm]{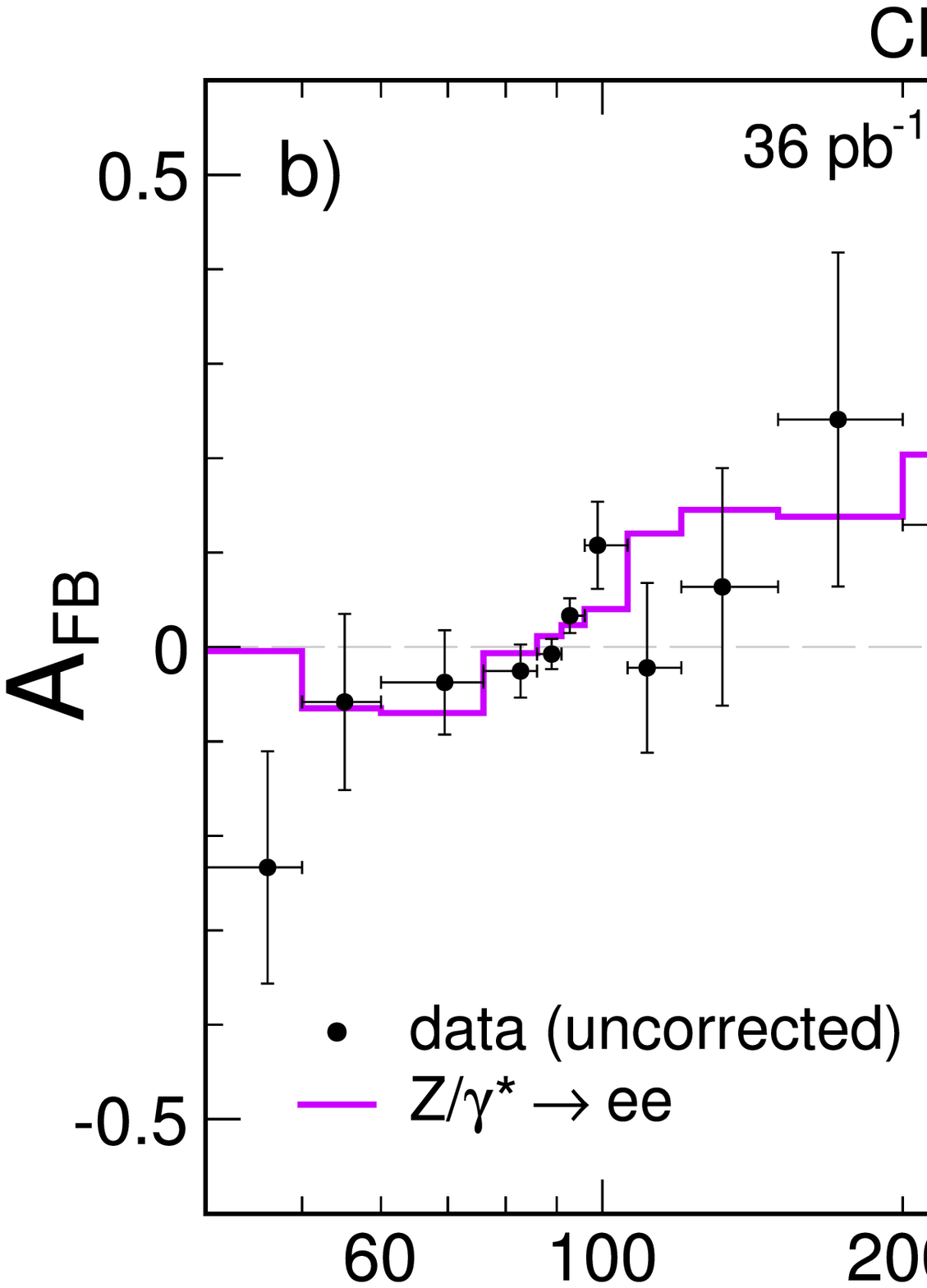}
\caption{Forward-backward asymmetry in electron channel. Top: D0 (corrected for detector effects). Bottom: CMS (uncorrected).}
\label{fig:afbd0cms}
\end{figure}

Measured forward-backward asymmetry in electron channel
 as a function of $Z$ boson mass from D0~\cite{d0afb} and CMS~\cite{cmsafb} is shown in Figure~\ref{fig:afbd0cms}.
CMS also measured forward-backward asymmetry in the muon channel.
Asymmetry is more pronounced in case of D0 than CMS mainly due to the nature of underlying proton-antiproton versus
proton-proton collisions at Tevatron and LHC. In case of Tevatron directions of the colliding quark and anti-quark are known since
they are valence quark constituents of proton and antiproton whose directions are known. In case of LHC $Z$ boson is produced
in a collision of a valence quark and sea anti-quark whose directions are unknown a priori. At LHC classification of events into forward and backward
is based on the sign of rapidity of the $Z$ boson. 

\subsection{Effective weak mixing angle	     }
\begin{figure}
\includegraphics[width=75mm]{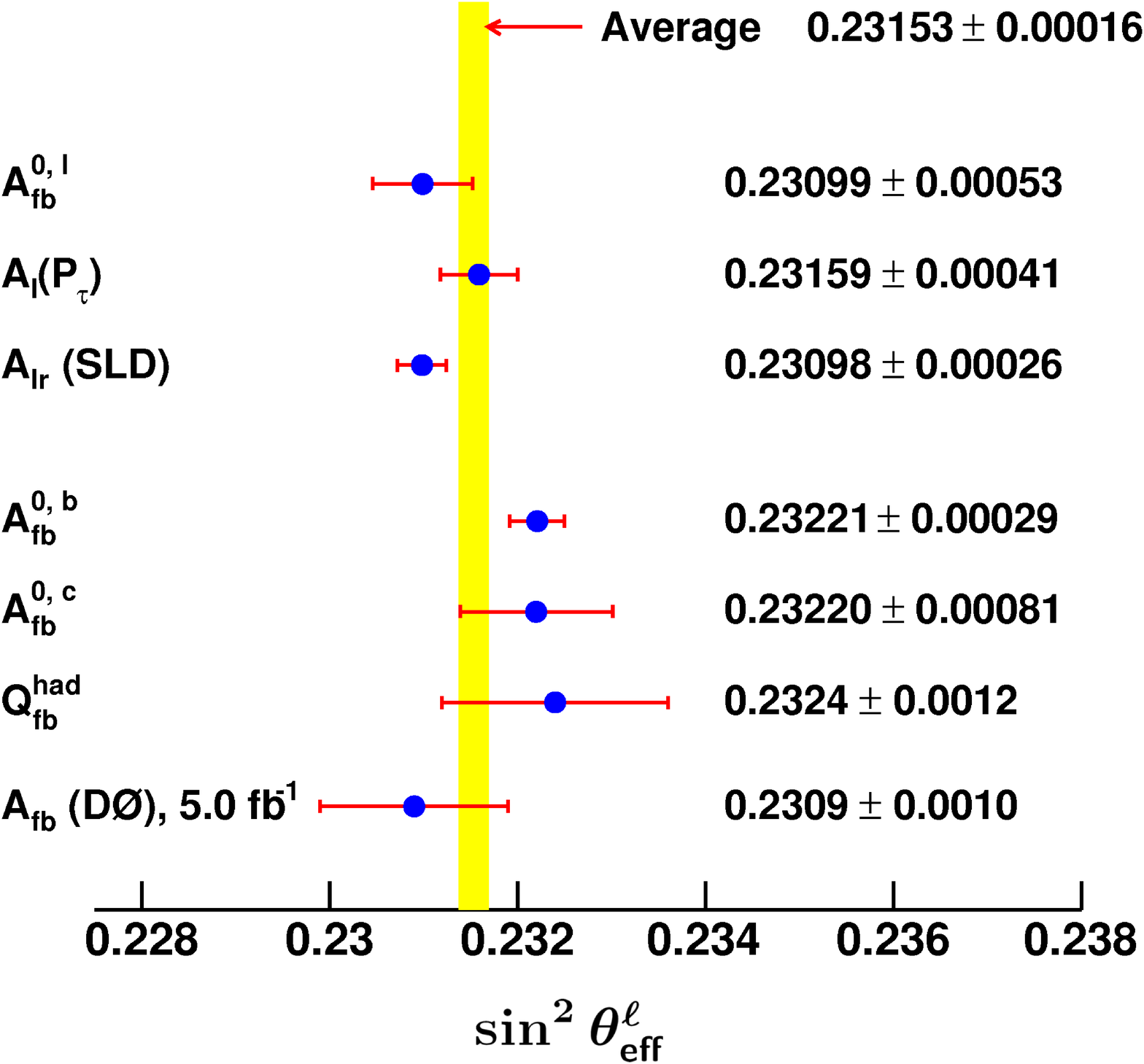}
\caption{Measurements of $sin^{2}\theta^{l}_{eff}$}
\label{fig:anglecomp}
\end{figure}

Using template fit to forward-backward asymmetry D0  measured effective weak mixing angle with 5.0fb$^{-1}$~\cite{d0afb}.
D0 result is $sin^{2}\theta^{l}_{eff}$ = 0.2306 $\pm$ 0.0010.
CMS used multivariate analysis of dilepton mass, rapidity and decay angle to measure effective weak mixing angle. 
CMS result with 234pb$^{-1}$ is $sin^{2}\theta^{l}_{eff}$ = 0.2287 $\pm$ 0.0020(stat.) $\pm$ 0.0025(syst.)~\cite{cmcweakangle}.
Figure~\ref{fig:anglecomp} shows comparisons of current measurements of $sin^{2}\theta^{l}_{eff}$ from different experiments.
CDF converted 2.1fb$^{-1}$ measurement of angular coefficient $A_4$~\cite{cdfangular} (see next section)
into measurement of effective weak mixing angle ~\cite{cdfpubweak}. 
CDF result is $sin^{2}\theta^{l}_{eff}$ = 0.2329 $\pm$ 0.0008($A_4$ error) $(\mbox{QCD})^{+ 0.0010}_{- 0.0009}$.

\subsection{Couplings between $\bf{Z}$ boson and fermions}
\begin{figure}
\includegraphics[width=75mm]{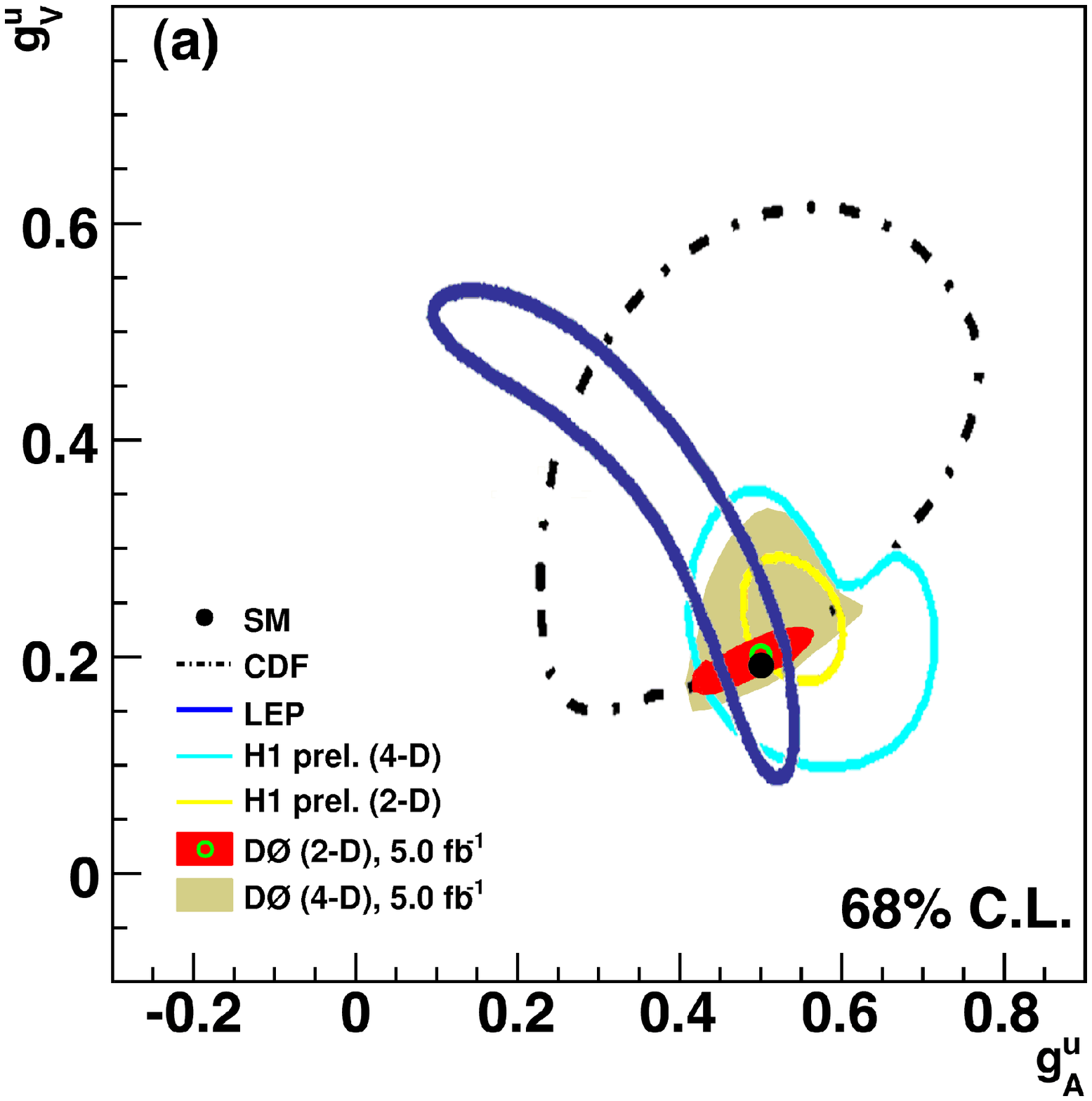}
\includegraphics[width=75mm]{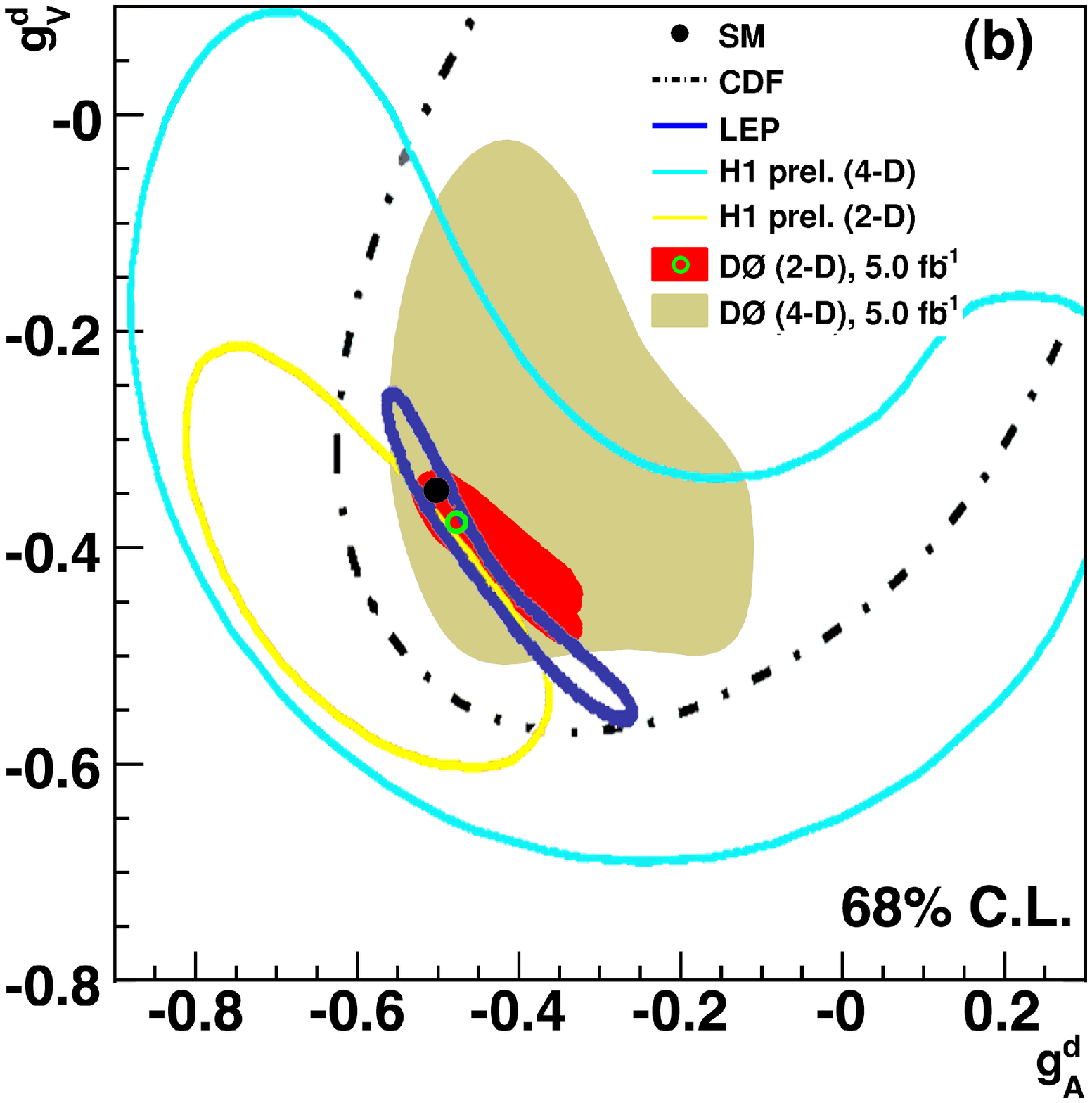}
\caption{Measurements of vector and axial-vector couplings of $u$ and $d$ quarks to the $Z$ boson}
\label{fig:udcouplings}
\end{figure}

D0 performed most precise direct measurements of the vector and axial-vector couplings of $u$ and $d$ quarks to the $Z$ boson~\cite{d0afb}.
These measurements are shown as two-dimensional contours and compared with measurements from other experiments in Figure~\ref{fig:udcouplings}.

\subsection{Angular coefficients }
\begin{figure}
\includegraphics[width=75mm]{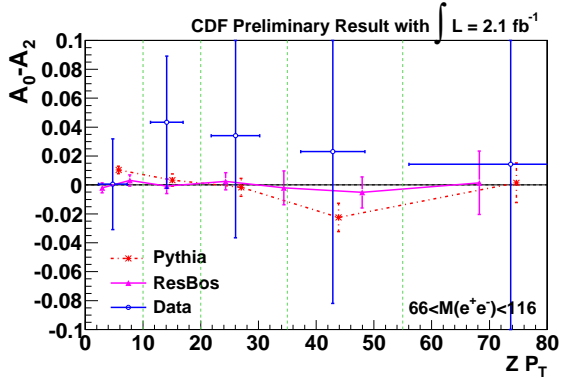}
\caption{Lam-Tung relation ( $A_{0} - A_{2}$ = 0 ) in $Z$ boson $p_T$ measurement (CDF).}
\label{fig:lamtung}
\end{figure}

CDF studied angular distributions of final state electrons using 2.1fb$^{-1}$ $Z \to ee + X$ sample~\cite{cdfangular}.
Angular distributions are described in Collins-Soper frame~\cite{csframe} in terms of polar and azimuthal angles~\cite{zangular}.
\begin{eqnarray*}
 d\sigma \over dcos\theta & \propto & (1 + cos^{2}\theta) + {1 \over 2} A_{0} (1 - 3cos^{2}\theta) + A_{4}cos\theta \\
  d\sigma \over d\phi     & \propto & {3 \pi A_{3} \over 16} cos\phi +  {A_{2} \over 4} cos2\phi
\end{eqnarray*}
Measurements of angular coefficients as a function of $Z$ boson transverse momentum are unique probes of underlying QCD production mechanism.
In particular, verification of Lam-Tung relation ($A_{0} - A_{2}$ = 0) is an indirect measurement of the spin of the gluon.
Lam-Tung relation is valid if gluon spin is 1 and badly broken if gluon spin is 0. 
Figure~\ref{fig:lamtung} shows measured $A_{0} - A_{2}$ as a function of $Z$ boson transverse momentum.
Also from $A_{0}$ and $A_{2}$ measurements 
relative contribution of annihilation and Compton production mechanisms was studied. It was concluded that at high $Z$ boson $p_T$
contribution from both processes is significant.

\section{$\bf{W}$ BOSON PROPERTIES}
Measurements of the following $W$ boson properties are presented:
\begin{enumerate}
\item $W$ polarization.
\item $W$ charge asymmetry.
\item $W$ mass and width.
\end{enumerate}

\subsection{$\bf{W}$ polarization}
Using $W$ boson decays in electron and muon channel CMS measured polarization
fractions $f_L$, $f_R$, and $f_0$~\cite{cmspolariz}. The results are found to be
consistent with SM prediction. 
The results are ($f_L$ - $f_R$)- = 0.226 $\pm$ 0.031 (stat.) $\pm$ 0.050 (syst.)
 and $f_0$- = 0.162 $\pm$ 0.078 (stat.) $\pm$ 0.136 (syst.) for negatively charged W bosons 
and ($f_L$ - $f_R$)+ = 0.300 $\pm$ 0.031 (stat.) $\pm$ 0.034 (syst.) 
and $f_0$+ = 0.192 $\pm$ 0.075 (stat.) $\pm$ 0.089 (syst.) for positively charged W bosons. 
This establishes for the first time that W bosons produced 
in pp collisions are predominantly left-handed, as predicted by the Standard Model.


\subsection{$\bf{W}$ charge asymmetry}
$W$ charge asymmetry arises from the asymmetric momentum fraction distributions
of the colliding partons which produce $W$ boson.
In case of proton-antiproton collisions $W^+(W^-)$ is produced with valence $u$ and $\bar{d}$
($d$ and $\bar{u}$) quarks.
As the $u$ quark tends to carry a higher fraction of the proton's 
momentum than the $d$ quark, the $W^+(W^-)$ is boosted, on average, 
in the proton(anti-proton) direction. 
Hence asymmetry in the production rate between $W^+$ and $W^-$ as a function of $W$ rapidity is observed.
Total number of produced $W^+$ and $W^-$ is the same.
In case of proton-proton collisions $W^+(W^-)$ 
is produced with valence $u$ and sea $\bar{d}$ quarks
(valence $d$ and sea $\bar{u}$) quarks. Qualitatively the same type of asymmetry as
a function of $W$ boson rapidity is expected
as in case of proton-antiproton collisions. However the shape is expected to differ
compared to proton-antiproton collisions
since valence quarks and sea quarks have different momentum fraction distributions.
Besides the total number of produced $W^+$ is expected to exceed that of $W^-$
since proton contains two valence $u$ quarks and one valence $d$ quark.
The inclusive ratio of cross sections for $W^+$ and $W^-$ boson production
was measured by CMS to be 1.43$\pm$0.05~\cite{cmswasymold}.

Since $W$ charge asymmetry is driven by parton distributions, by measuring the asymmetry
parton distribution functions can be constrained. Central $W$ events correspond to energetically
symmetric collisions. In contrast, when one of the colliding quarks carries large
fraction of proton momentum and the other quark carries small fraction, 
$W$ bosons are produced at large rapidities. Therefore, the wider rapidty range of the asymmetry measurements
the wider range of momentum fractions can be constrained.

Asymmetry in the $W$ boson rapidity distribution
has traditionally been studied in terms of 
charged lepton asymmetry, as $W$ boson rapidity cannot be determined
on the event-by-event basis, since neutrino escapes the detection.
Charged lepton asymmetry, is the convolution of $W^{\pm}$ production 
and V-A (vector-axial vector) decay asymmetries.
Asymmetry is defined as a ratio of difference and sum of positively charged and negatively charged leptons.

\begin{figure}[hbp]
\begin{center}
\includegraphics [width=75mm] {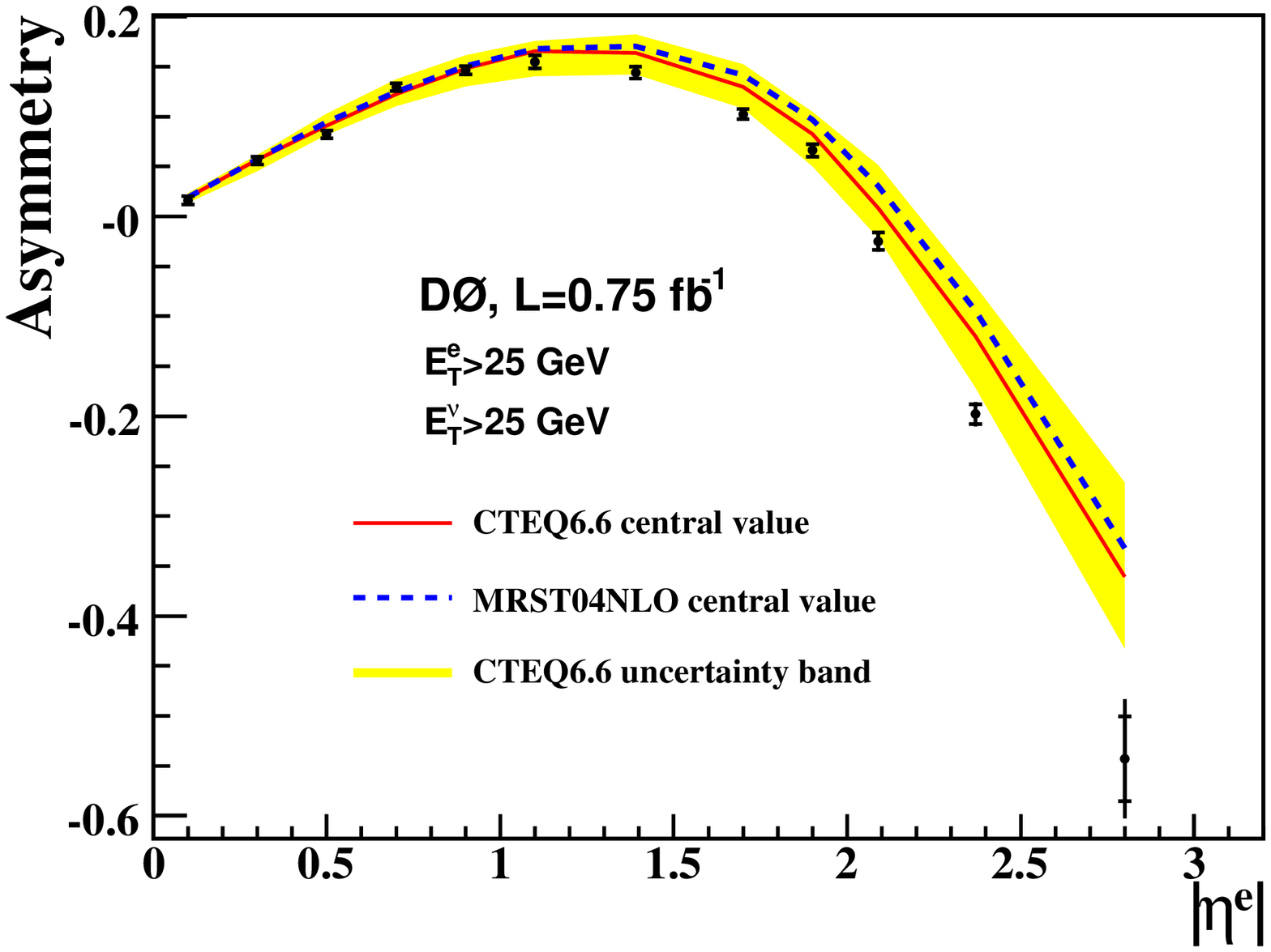}
\includegraphics [width=75mm] {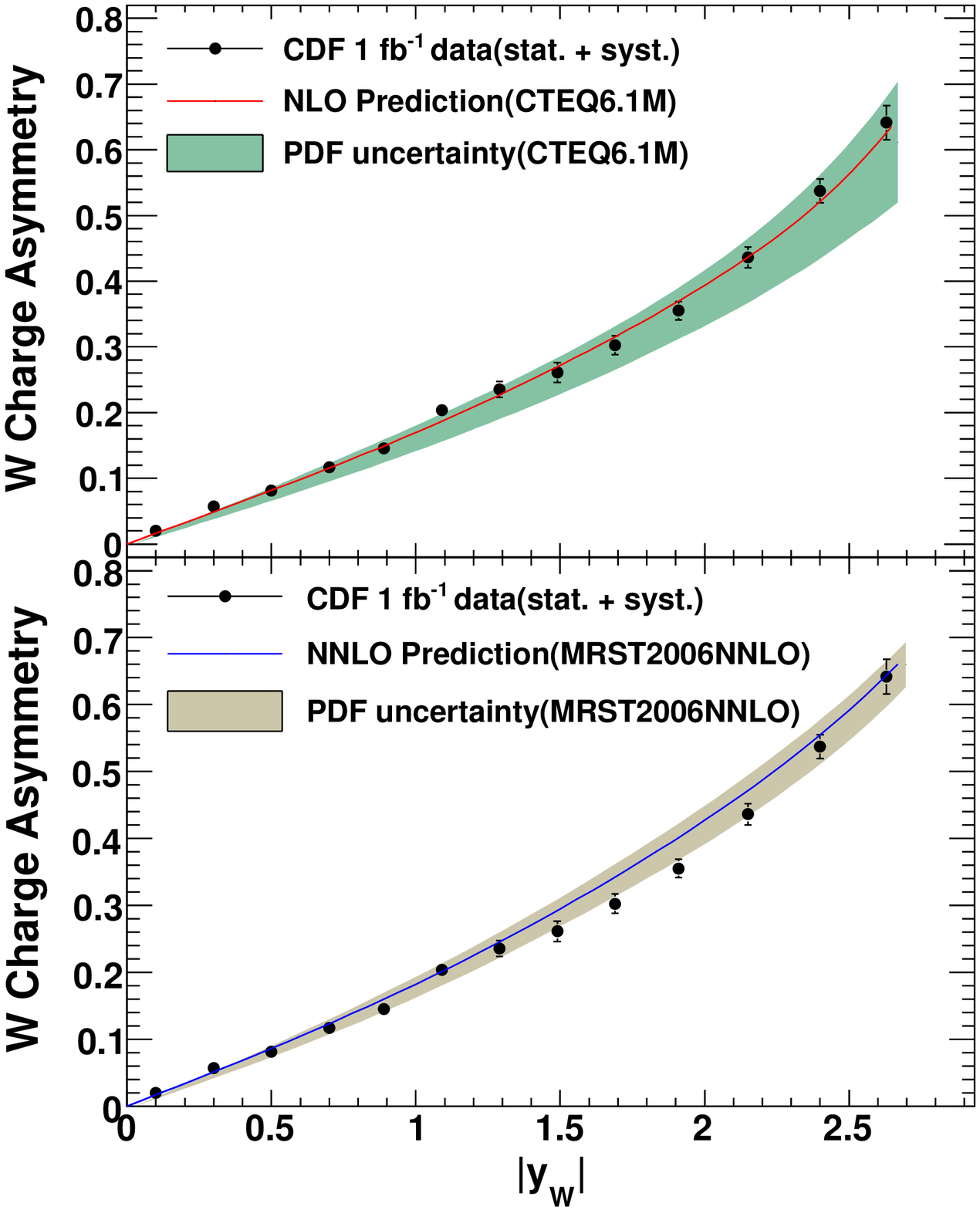}
\end{center}
\vspace{-1pc}
\caption{
Top: electron charge asymmetry (D0). 
Middle and bottom: $W$ boson charge asymmetry (CDF)} 
\label{fig:wasym_D0CDF}
\end{figure}

\begin{figure}[hbp]
\begin{center}
\includegraphics [width=75mm] {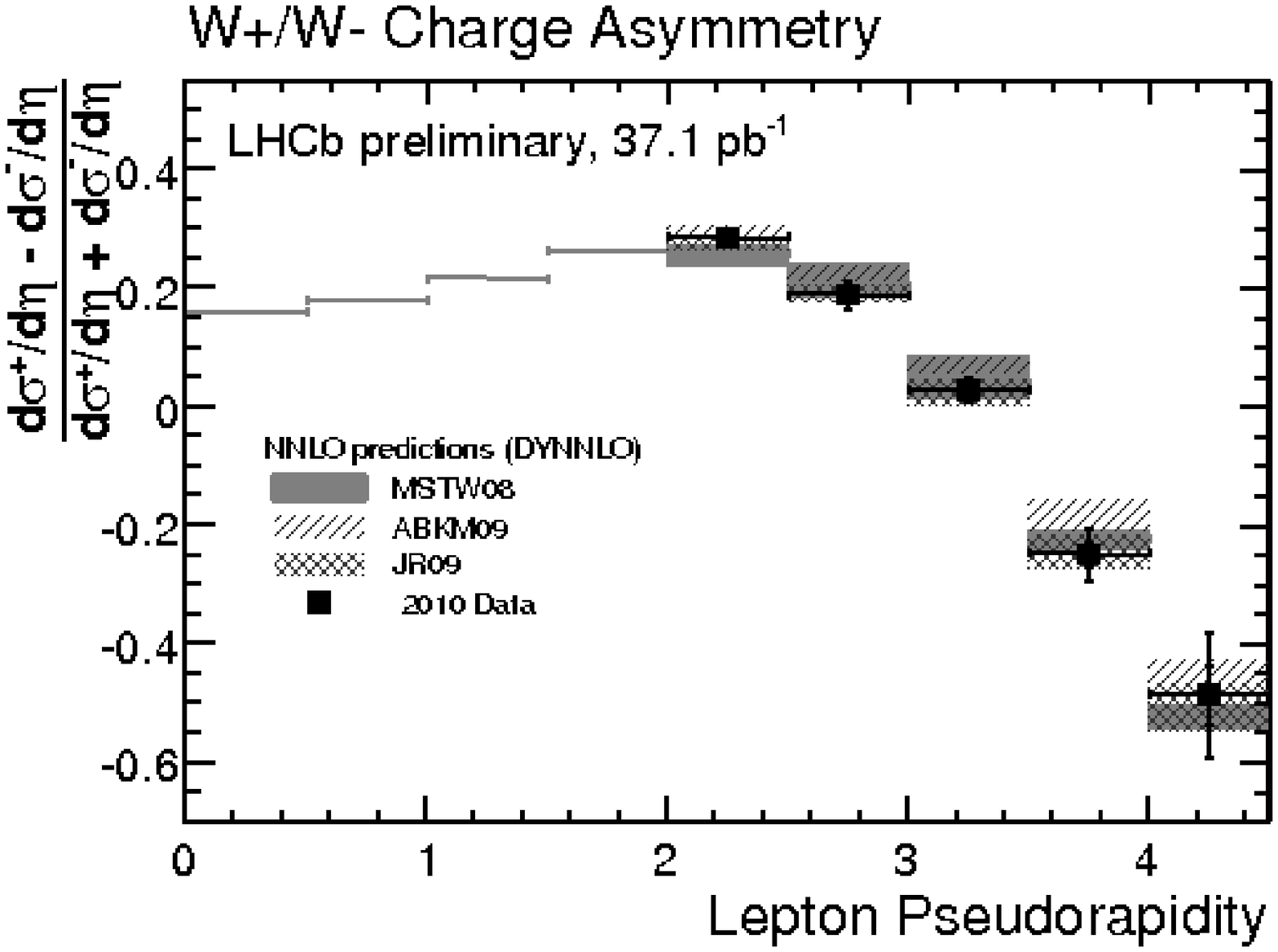}
\includegraphics [width=75mm] {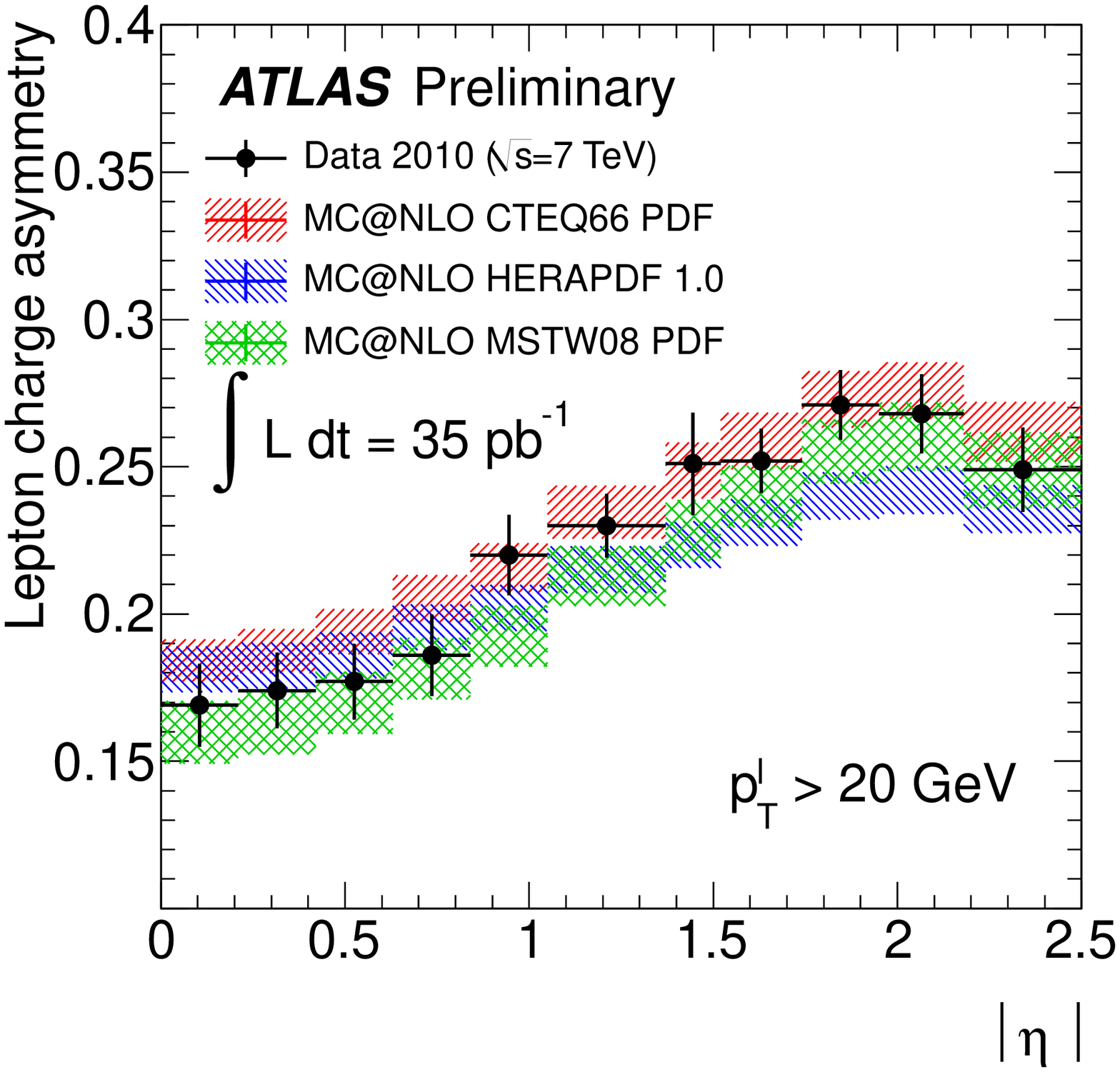}
\includegraphics [width=75mm] {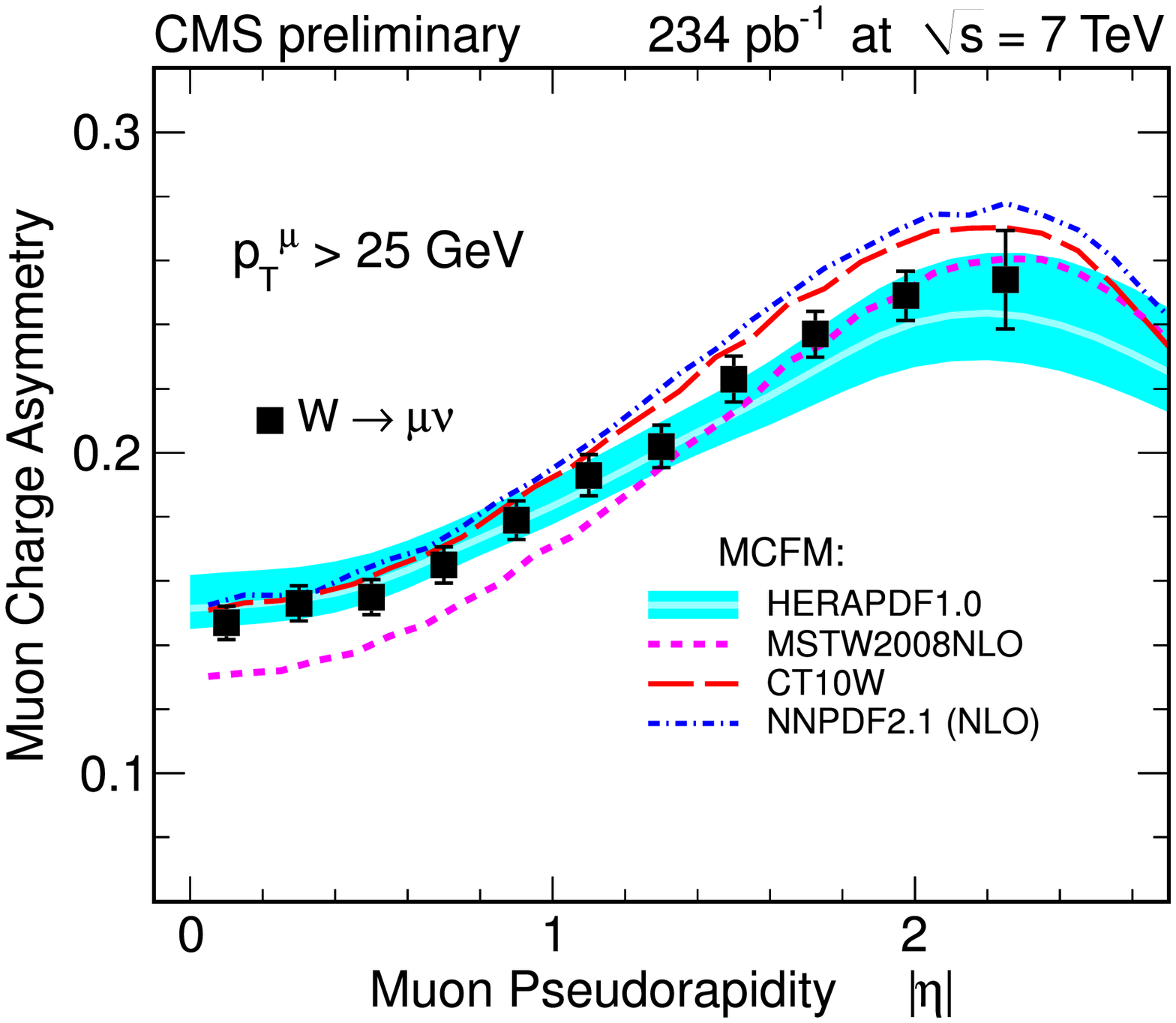}
\end{center}
\vspace{-1pc}
\caption{Lepton charge asymmetry at LHC. Top: LHCb~\cite{lhcbswasym}. Middle: ATLAS~\cite{atlaswasym}. Bottom: CMS~\cite{cmswasym}.} 
\label{fig:wasym_LHC_indiv}
\end{figure}

\begin{figure}[hbp]
\begin{center}
\includegraphics [width=75mm] {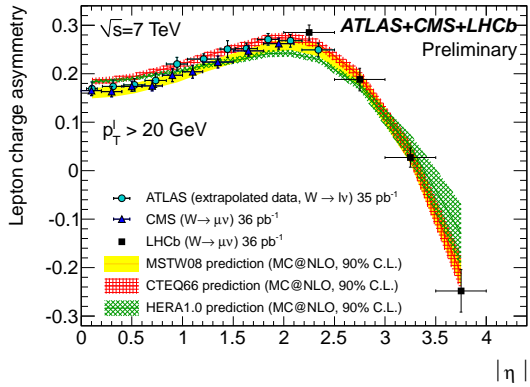}
\end{center}
\vspace{-1pc}
\caption{ATLAS+CMS+LHCb combined measurement of lepton charge asymmetry} 
\label{fig:wasym_LHC_comb}
\end{figure}

D0 electron charge asymmetry distribution~\cite{d0elasym} is shown on the top of Figure~\ref{fig:wasym_D0CDF}.
CDF used a sophisticated method which allows to measure
$W$ asymmetry~\cite{cdfwasym} rather than lepton asymmetry. In this method neutrino
momentum is determined using $W$ mass constraint, $W$ decay structure,
and $W$ production cross-section as a function of rapidity.
Measured $W$ charge asymmetry and comparisons with theory are shown
in  Figure~\ref{fig:wasym_D0CDF}.
Figure~\ref{fig:wasym_LHC_indiv} shows measured lepton charge asymmetries and comparisons with theory at LHC.
Figure~\ref{fig:wasym_LHC_comb} shows the corresponding ATLAS+CMS+LHCb combined measurement.
\clearpage

\subsection{W mass and width}
A precision measurement of $M_{W}$ is one of the highest priorities for the Tevatron experiments.
$M_{W}$ measurement combined with precise measurement of the top quark mass ({$M_{top}$), 
constrains the mass of the Higgs boson.
Most precise measurement of $M_{W}$ by the D0 collaboration~\cite{d0mw} in the $W \rightarrow e \nu$  
decay mode with an integrated luminosity of 1$fb^{-1}$ along with
other $M_W$ measurements and combinations is shown in Figure~\ref{fig:mwfig}

$\Gamma_{W}$ is an important parameter of SM. D0 measurement
$\Gamma_{W} = 2.028 \pm 0.039 \textrm{(stat)} \pm 0.061 \textrm{(syst)}$ GeV is the most 
precise measurement of this quantity to date. 
\begin{figure}[hbpt]
\begin{center}
  \includegraphics [width=75mm] {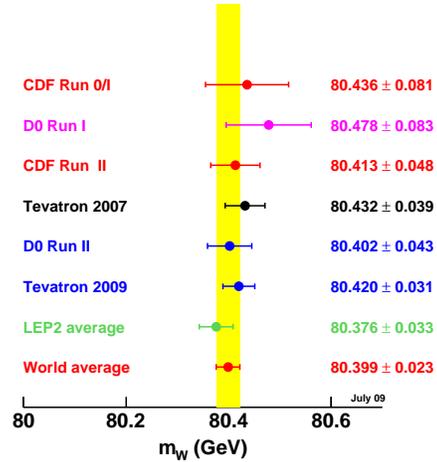}
 \end{center}
\vspace{-1pc}
  \caption{
 Summary of the measurements of the $W$ boson mass and their average. 
The result from the Tevatron corresponds to the values which includes 
corrections to the same W boson width and PDFs. 
The LEP II results are from~\cite{lep}. 
An estimate of the world average of the Tevatron and LEP 
results is made assuming no correlations between the Tevatron and LEP 
uncertainties.
}
\label{fig:mwfig}
\end{figure}

\section{CONCLUSION}
A number of measurements of electroweak gauge boson properties from both the Tevatron and the LHC 
experiments were presented. These include limits on anomalous couplings between the gauge bosons,
couplings between Z boson and fermions, measurements of forward-backward asymmetry, effective
weak mixing angle, angualar coefficients, as well as charge asymmetry, mass, width and polarization of W boson.

\clearpage

%

\end{document}